\documentclass[epj,referee]{svjour}
\usepackage{graphics}

\begin{document}

\title{Thermal critical behavior and universality aspects of the three-dimensional random-field Ising model}
\author{A. Malakis\thanks{e-mail: amalakis@phys.uoa.gr} \and N.G. Fytas\thanks{e-mail: nfytas@phys.uoa.gr}
}
\institute{Department of Physics, Section of Solid State Physics,
University of Athens, Panepistimiopolis, GR 15784 Zografos,
Athens, Greece}

\date{Received: date / Revised version: date}

\abstract{The three-dimensional bimodal random-field Ising model
is investigated using the N-fold version of the Wang-Landau
algorithm. The essential energy subspaces are determined by the
recently developed critical minimum energy subspace technique, and
two implementations of this scheme are utilized. The random fields
are obtained from a bimodal discrete $(\pm\Delta)$ distribution,
and we study the model for various values of the disorder strength
$\Delta$, $\Delta=0.5,\;1,\;1.5$ and $2$, on cubic lattices with
linear sizes $L=4-24$. We extract information for the probability
distributions of the specific heat peaks over samples of random
fields. This permits us to obtain the phase diagram and present
the finite-size behavior of the specific heat. The question of
saturation of the specific heat is re-examined and it is shown
that the open problem of universality for the random-field Ising
model is strongly influenced by the lack of self-averaging of the
model. This property appears to be substantially depended on the
disorder strength.
}
\authorrunning{A. Malakis and N.G. Fytas} \titlerunning{Thermal critical behavior and
universality aspects of the three-dimensional...}

\maketitle

\section{Introduction}
\label{intro}

The random-field Ising model (RFIM)~\cite{imry75} is one of the
most studied glassy magnetic
models~\cite{natter88,belang91,rieger95a,belang98}, mainly because
of its interest as a simple frustrated system. The Hamiltonian of
the system is:
\begin{equation}
\label{eq:1}
\mathcal{H}=-J\sum_{<i,j>}S_{i}S_{j}-\sum_{i}h_{i}S_{i},
\end{equation}
where the $S_{i}=\pm1$ are Ising spins, $J$ is the interaction
energy between nearest neighbors, which we take to be positive so
that the model is ferromagnetic and $h_{i}$ are the random fields
(RF's). In this paper the values $h_{i}$ are taken from a bimodal
distribution of the form:
\begin{equation}
\label{eq:2} P(h_{i})=\frac{1}{2}
\delta(h_{i}-\Delta)+\frac{1}{2}\delta(h_{i}+\Delta),
\end{equation}
with $\Delta$ the disorder strength, also called randomness of the
system. Various different RF probability distributions have been
studied in the
past~\cite{hartmann01,hartmann99,newman96,sourlas99}, such as the
Gaussian distribution, the wide bimodal distribution (with a
Gaussian width), and the bimodal distribution considered also
here.

In spite of many years of study, the critical behavior of the
three-dimensional (3D) RFIM has been a matter of several debates
and is still controversial. One of the early disagreements was the
question of whether the model undergoes a phase transition from a
high temperature paramagnetic phase to a low temperature
ferromagnetic one, for some range of the randomness $\Delta$. The
work of Parisi and Sourlas~\cite{parisi78} introduced the notion
of dimensional reduction, indicating that the critical behavior of
the RFIM in $d$ dimensions, at sufficiently low randomness, should
be identical to that of the well-known normal Ising model in $d-2$
dimensions. This in turn indicated that the model should not
exhibit a phase transition in 3D or fewer. However, a different
argument based on the droplet theory of domain wall energies in
the ferromagnetic state~\cite{grinstein82}, suggested that a phase
transition should exist in 3D for finite temperature and
randomness. The whole puzzle has been largely cleared out by
Imbrie~\cite{imbrie84} and Bricmont and
Kupiainen~\cite{bricmont87}, who showed the existence of an
ordered phase. Their arguments strongly supported the view that a
phase transition in 3D exists, provided that the randomness is
sufficiently small $(\Delta_{c}\simeq2.3)$.

However, agreement over several fundamental issues is missing and
the characterization of the phase transition is still
unclear~\cite{rieger95}. Despite the fact that most studies
suggest a second-order
transition~\cite{rieger95,villain85,fisher85,bray85,ogielski86,ogielski86b},
there are also indications of first-order or hybrid-order
transition~\cite{rieger95,mccay88}. Note also that the mean field
theory~\cite{aharony78} differentiates between a binary and a
continuous randomness distribution, predicting for the former a
tricritical point at which the transition becomes of the
first-order, at high fields. However, it is now generally accepted
that a new fixed point controls the behavior of RF
ferromagnets~\cite{chayes86,cardy}. The significance of this for
the RFIM (in $d>2$) is that this new zero temperature random fixed
point controls the whole critical line ($T_{c}(\Delta)$) and that
the RF's are always relevant. For disordered systems with weak
randomness which couples to the local energy (such as random-site
impurity or random-bond models) the crossover to a new random
fixed point, depends on the Harris
criterion~\cite{cardy,harris74}. According to this, the disorder
is relevant if the correlation length exponent of the pure model
($\nu=\nu_{pure}$) satisfies the condition $d\nu<2$ and this
condition may be stated, with the help of the hyper-scaling
relation ($\alpha=2-d\nu$), as $\alpha>0$. Since the specific heat
exponent of the 3D Ising model is positive, weak disorder should
be expected to be relevant. In the case of the RFIM the type of
disorder is much more severe, since the randomness couples to the
local order-parameter and the crossover renormalization group
eigenvalue is always positive~\cite{cardy}. The inequality
$\nu\geq 2/d$, derived by Chayes \emph{et al.}~\cite{chayes86} for
the correlation length exponent of a generic disordered system
($\nu=\nu_{random}$) would imply, using again hyper-scaling, a
negative specific heat exponent ($\alpha<0$). However, it is
believed that hyper-scaling is violated in the RFIM and the
specific heat exponent $\alpha$ is related to $\nu$ by a modified
hyper-scaling law $2-\alpha=(d-\theta)\nu$, where the exponent
$\theta$ characterizes the scaling of the stiffness of the ordered
phase at the critical point. Thus, the specific heat exponent of
the RFIM is not restricted, by the above theoretical
considerations, to be negative~\cite{chayes86}.

A general sketch of the phase diagram of the RFIM is given in
several papers~\cite{hartmann01,newman96,itakura01} and will be
also presented here in Sect.~\ref{sec:3c}. At low temperatures and
moderate values of randomness, the system is assumed to be in an
ordered ferromagnetic phase, whereas in the opposite regime the
system is paramagnetic. From the notions of the perturbative
renormalization group (PRG) it is expected that the RF is the
relevant perturbation at the pure $(\Delta=0)$ fixed point, and
that the RF fixed point is at $T=0$. However, it is known that PRG
fails for the RFIM and that a theoretical justification of
universality for this and also other disordered systems is
lacking~\cite{natter88,belang91,sourlas99,parisi78}. Questions
concerning the general characterization of the phase transition,
the existence of an intermediate glassy
phase~\cite{itakura01,mezard92,mezard94}, the behavior of the
renormalization group flow in the middle of the phase
diagram~\cite{itakura01}, and the dependence of the critical
exponents on the randomness distribution and disorder strength are
still open~\cite{hartmann99,sourlas99,sourlas97}.

A relevant active and enigmatic issue concerns the behavior of the
specific heat (see Ref.~\cite{fytas06a} and references therein).
The specific heat of the RFIM can be experimentally measured and
is of considerable theoretical interest. There is a strong
disagreement in literature about the possible divergence or
saturation of the specific heat. In studies supporting the
scenario of saturation there is a discrepancy in the reported
negative values of the critical exponent $\alpha$. Some of these
later studies find strongly negative values, ranging from
$\alpha=-1.5$~\cite{rieger93} to
$\alpha=-0.5$~\cite{hartmann01,rieger95,nowak98}. In particular,
Hartmann and Young~\cite{hartmann01} recently found by a ground
state technique the value $\alpha=-0.63\pm 0.07$, whereas
Middleton and Fisher~\cite{middle02}, using the same technique,
estimated in marked disagreement $\alpha=-0.01\pm0.09$.

From the experimental point of view, a true realization of the
RFIM is hardly conceived. However, it has been shown that dilute
antiferromagnets in uniform external field (DAFF) represent
physical realizations of the RFIM~\cite{fishman79} and a number of
experiments investigated the phase transitions of such 3D
systems~\cite{belanger83}. These experiments have proven to be
very difficult and their interpretation doubtful due to the
extremely slow, glassy dynamics of the system. Experiments on
DAFF, provided evidence of a second-order phase transition and a
logarithmic singularity for the specific heat~\cite{belanger85}.
Note that recently, Barber and Belanger~\cite{barber01} in their
Monte Carlo study of a DAFF model reported also that their
specific heat curve closely mimics a logarithmic peak. Their
results are based on a large lattice $(L=128)$ but instead of
sample averaging they have observed the behavior of only a few RF
realizations. On the other hand, there is also experimental
evidence~\cite{karszewiski94} supporting the opposite view of a
cusp-like singularity of the specific heat, in agreement with a
saturating specific heat $(\alpha<0)$ as found in the studies of
Refs.~\cite{rieger95,rieger93}.

It has been pointed out that a strongly negative value of $\alpha$
causes serious difficulties with respect to the Rushbrooke
relation:
$\alpha+2\beta+\gamma\geq2$~\cite{hartmann01,rieger95,nowak98,middle02}.
Therefore, there have been several
attempts~\cite{hartmann01,villain85,fisher85,bray85,grinstein76}
in order to find a consistent set of scaling relations to describe
the critical behavior of the RFIM. Among the several scaling
scenarios
proposed~\cite{sourlas99,mezard92,mezard94,sourlas97,nowak98,middle02},
the single second-order critical point behavior characterized by
three scaling exponents~\cite{middle02} seems to be consistent
with a close to zero estimate for the specific heat exponent.
Thus, the above described conflicting situation in literature
concerning the divergence or saturation of the specific heat is
one of the open important issues, whose implications on the
critical behavior of the model are not understood.

The first step towards its resolution was taken recently by the
present authors~\cite{fytas06a,fytas06b}, where an extended
numerical investigation of the 3D bimodal ($\Delta=2$) RFIM
revealed the importance of the property of lack of self-averaging
of the specific heat of the model, as well as the possibility of
large-$L$ crossover phenomena in the scaling behavior of the
specific heat of the model. Here, we extend our analysis for the
values $\Delta=0.5,\;1$ and $2$ of the disorder strength below the
critical value $\Delta_{c}$ in order to obtain a more
comprehensive picture. To this end, we implement recently
developed efficient Monte Carlo methods that directly calculate
the density of states (DOS) of a classical statistical model. A
brief overview of the numerical techniques used in the past for
the RFIM are presented in the next Section, together with the
necessary details of the methods used in our approach. The
utilization of our recently proposed critical minimum energy
subspace (CrMES) scheme~\cite{malak04} to the RFIM is also
explained. In Sect.~\ref{sec:3} the new numerical results for the
cases $\Delta=0.5,\;1$ and $\Delta=1.5$ are given and the phase
diagram of the model is reproduced. The universality aspects of
the model are discussed and found to support the scenario of
violation of universality. Our conclusions are summarized in
Sect.~\ref{sec:4}.

\section{Numerical techniques}
\label{sec:2}

There are two distinct kinds of numerical approaches for the RFIM.
In the first approach, traditional Monte Carlo methods are used to
simulate the properties of the system at finite
temperatures~\cite{rieger95,ogielski86,rieger93,nowak98,barber01,metrop53,newman99,young86}.
The second approach is grounded on the well-known belief that the
critical behavior of the model is governed by the non trivial RF
fixed point at $T=0$~\cite{middle02,dukov03}. In this case, graph
theoretical algorithms~\cite{hartmann99,swamy91} are used to
calculate the ground states of the system for a sample of RF's at
different disorder strength. Using this later approach quite large
lattices have been studied: $L\leq80$~\cite{hartmann99,dukov03},
$L\leq90$~\cite{sourlas99}, $L\leq96$~\cite{hartmann01} and
finally $L\leq256$~\cite{middle02}. Yet, in the traditional Monte
Carlo approach the sizes studied were restricted to the size
$L\leq16$~\cite{newman96,rieger95,rieger93},
$L\leq20$~\cite{nowak98} and finally we may refer, as an
exception, to the case $L=128$ in the study of
Ref.~\cite{barber01} for particular RF's as mentioned in the
introduction. From the $T=0$ numerical studies one obtains an
accurate estimate of the critical randomness and from the finite
temperature studies further information for the phase diagram may
be derived. From the finite temperature approach one can also
find, by extrapolation, a crude estimate for the critical
randomness~\cite{newman96} (see Sect.~\ref{sec:3}). It is worth
noting that quite recently Wu and Machta, combining finite and
zero temperature studies of the RFIM, reported strong correlations
of ground states and thermal states near the critical line for
given realizations of the disorder, supporting strongly the $T=0$
fixed point scenario~\cite{wu05}.

In traditional Monte Carlo studies of the
RFIM~\cite{newman96,rieger93,ogielski86,rieger95,nowak98} the
system is simulated in a restricted range of temperatures,
appropriate for the location of the pseudocritical temperatures.
However, single spin-flip methods, such as the Metropolis or the
heat bath algorithms, face severe slowing down problems of
equilibration and temperature averaging since the characteristic
times may be exponentially large at low temperatures $(T<T_{c})$,
as explained in Ref.~\cite{newman96}. Moreover, the sample
averaging process introduces new characteristic features and
requires further computer resources. Indeed, the appropriate
pseudocritical temperature for the RFIM is a strongly fluctuating
quantity~\cite{fytas06a}, and this property amplifies the computer
time requirements for its location. Hence, depending on the size
of the lattice and the disorder strength, it is necessary to
simulate the system for each RF realization in a quite wide
temperature range, which is not even known in advance. To obtain a
good approximation of the locations of the specific heat peaks,
the temperature step must be chosen sufficiently small for,
otherwise any interpolation scheme may miss the correct height of
a possible sharp peak. In fact, this situation of a possible sharp
peak, turns out for a significant number of RF's~\cite{fytas06a}.

From the above discussion one should wonder whether the
traditional Metropolis sampling could be trusted to provide even a
moderate estimation plan, since it requires immense computer
resources and faces all mentioned problems. The cluster flipping
algorithm for the RFIM proposed by Dotsenko, Selke and
Talapov~\cite{dotsenko91} is a straightforward extension of the
Wolff algorithm~\cite{wolff89}, devised to overcome the slowing
down effect and speed up the flip dynamics. A more efficient form
of this algorithm, the limited cluster flip (LCF) algorithm, has
been invented by Newman and Barkema~\cite{newman96} and was used
for the study of the Gaussian RFIM. Furthermore, these authors
have combined the LCF algorithm with the generalized histogram
method of Ferrenberg and Swendsen~\cite{ferre98} which is a
re-weighting scheme, using a restricted set of temperature
measurements. This combination may be hopefully more reliable for
the location of the pseudocritical temperatures. Finally, a new
cluster technique, that combines the replica-exchange method of
Swendsen and Wang~\cite{swend86} and the two-replica cluster
method~\cite{redner98}, was implemented by Machta, Newman, and
Chayes~\cite{machta00} where single realizations of the disorder
strength were studied for sizes up to $L=24^{3}$.

Here, we employ a different strategy which utilizes the new and
popular methods of efficient estimation of spectral degeneracies
of classical statistical
models~\cite{newman99,ferre98,malak04b,lee93,berg92,lima00,wang01,schul01,oliv96,zhou05}
and the recently developed CrMES technique~\cite{malak04}. This
scheme has the merit of locating the pseudocritical temperatures
by determining the DOS in the proper energy subspace by using
simple algorithms in a unified implementation. Moreover, it avoids
all the above problems, speeding up the simulations. Specifically,
we use the multi-range Wang-Landau (WL) algorithm~\cite{wang01},
and its N-fold version as presented by Schulz \textit{et
al.}~\cite{schul01}. The accuracy of this scheme was discussed in
Ref.~\cite{fytas06a}, where more details than those given below
for the appropriate implementations can be found.

\subsection{The Wang-Landau algorithm}
\label{sec:2a}

For the application of the WL algorithm in a multi-range approach
we follow the description of Schulz \textit{et
al.}~\cite{schul01}, i.e., whenever the energy range is restricted
we use the updating scheme $2$ in that paper. Consider the
restriction of the random walk in a particular energy range
$I=[E_{1},E_{2}]$ and assume that the random walk is at the border
of the range $I$. Then, the next spin-flip attempt is determined
by the modified Metropolis acceptance ratio:
\begin{equation}
\label{eq:3} A=\left\{\begin{array}{rr}
\min\{1,G(E)/G(E+\Delta E)\}, & \mbox{$(E+\Delta E)\in I$} \\
0, & \mbox{$(E+\Delta E)\not\in I$} \end{array}\right.,
\end{equation}
the random walk is not allowed to move outside of the energy
range, and we always increment the histogram $H(E)\rightarrow
H(E)+1$ and the DOS $G(E)\rightarrow G(E)*f_{j}$ after a spin-flip
trial. Here, of course, $f_{j}$ is the value of the WL
modification factor $f$~\cite{wang01} at the jth iteration, in the
process $(f\rightarrow f^{1/2})$ of reducing its value to 1, where
the detailed balance condition is satisfied. In all our
simulations the WL modification or control parameter takes the
initial value: $f_{j=1}=e\approx2.71828...$. When starting a new
iteration the control parameter is changed according to
$f_{j+1}=\sqrt{f_{j}}, j=1,2,\ldots,20$~\cite{wang01}. For the
histogram flatness criterion, we use a flatness level $0.05$, as
in previous studies~\cite{malak04,malak04b}.

\subsection{The N-fold version of the Wang-Landau algorithm}
\label{sec:2b}

For the bimodal RFIM is convenient to use an index $n$ to
characterize directly the corresponding energy changes produced by
the spin-flip process. The number of different classes for the
N-fold version $n=1,...,\mathcal{N}$ depends on the value of the
disorder strength $\Delta$. For example, consider the case
$\Delta=1$. The possible energy changes are $\Delta
E_{n}=\pm14,\;\pm10,\;\pm6$ and $\pm2$, and using an index
$n=1,2,...,8$ corresponding to $8$ classes we can write $\Delta
E_{n}=-14+(n-1)\cdot 4$. Note that, the RF value at the site in
which the spin-flip is going to take place is also affecting the
energy change. Denoting the populations of spins by $N_{n},
\sum_{n}N_{n}=N$, where $N$ is the number of sites: $N=L^{3}$, the
selection probability of a class, $P_{n}$, will be proportional to
this number multiplied by the corresponding acceptance ratio. For
the application of the algorithm in multi-range approach we follow
the description of Schulz \textit{et al.}~\cite{schul01} for the
N-fold version of the WL method. If the system is in a spin state
with energy $E\in I$ and after the spin-flip is in a state with
energy $E'=E+\Delta E_{n}$, the selection of the class for the
next spin-flip is obtained from the following~\cite{malak04b}:
\begin{equation}
\label{eq:4} P_{n}=N_{n}A_{n};\; A_{n}=\left\{\begin{array}{rr}
\min\{1,G(E)/G(E')\}, & \mbox{$E'\in I$} \\
0, & \mbox{$E'\not\in I$} \end{array}\right..
\end{equation}
The average life time of a state, reflecting the number of
attempts we expect the system to remain in its current state
before moving to the new state, is $\Delta t=W/Z$~\cite{schul01}
where:
\begin{equation}
\label{eq:5} W=\sum_{n}P_{n}.
\end{equation}
The rest of details for the algorithms can be found in the
original papers~\cite{malak04b,schul01}.

\subsection{Implementation of the CrMES technique}
\label{sec:2c}

The CrMES scheme~\cite{malak04} uses only a small part
$(\widetilde{E}_{-},\widetilde{E}_{+})$ of the energy space
$(E_{min},E_{max})$ to determine the specific heat peaks. If
$\widetilde{E}$ is the value of energy producing the maximum term
in the partition function at the temperature of interest (say the
pseudocritical temperature), the sums are restricted as follows:
\begin{eqnarray}
\label{eq:6}
C_{L}(\widetilde{E}_{-},\widetilde{E}_{+})=N^{-1}T^{-2}\left\{\widetilde{Z}^{-1}
\sum_{\widetilde{E}_{-}}^{\widetilde{E}_{+}}E^{2}\exp{[\widetilde{\Phi}(E)]}-\right.\nonumber\\\left.
\left(\widetilde{Z}^{-1}\sum_{\widetilde{E}_{-}}^{\widetilde{E}_{+}}E
\exp{[\widetilde{\Phi}(E)]}\right)^{2}\right\}
\end{eqnarray}
and
\begin{equation}
\label{eq:7} \widetilde{\Phi}(E)=[S(E)-\beta
E]-\left[S(\widetilde{E})-\beta
\widetilde{E}\right];\;\;\widetilde{Z}=\sum_{\widetilde{E}_{-}}^{\widetilde{E}_{+}}\exp{[\widetilde{\Phi}(E)]},
\end{equation}
where $(\widetilde{E}_{-},\widetilde{E}_{+})$ is the minimum
dominant subspace satisfying the following accuracy criterion:
\begin{equation}
\label{eq:8}
\left|\frac{C_{L}(\widetilde{E}_{-},\widetilde{E}_{+})}{C_{L}(E_{min},E_{max})}-1\right|\leq
r.
\end{equation}
In this paper we have used the accuracy criterion $r=10^{-4}$,
which is extremely demanding compared to the relative errors
produced in the specific heat, say by the WL method. It is also a
very strict criterion for the present model, in view of the
existing very large sample-to-sample fluctuations of the specific
heat. A practical algorithmic approach for specifying the CrMES is
fully described in Ref.~\cite{malak04}. We may satisfy the
specific heat accuracy criteria defined in equation~(\ref{eq:8})
for any particular realization of the RF, by restricting the WL
random walk in the corresponding critical energy subspace.

This restriction greatly facilitates our simulations without
introducing additional errors. Since we don't know in advance the
CrMES for a specific realization of the RF we have two
alternatives. The first option is to use an efficient prognostic
method of identifying the CrMES for any particular realization of
the RF by using the first stages of the WL method. For instance,
one may try to estimate the CrMES from the first $12$ iterations
in the process of reducing the WL modification factor $f$. To
implement safely this option, one should be careful to use for the
rest of WL iterations a much wider energy range than that
predicted in the first $12$ iterations. The second option is to
`guess' (by using some convenient extrapolation method) a broad
energy subspace that will cover the overlap of the CrMES for all
RF's in the sample. Implementing the second method is
straightforward and has the advantage that the approximation for
the specific heat curve of a particular RF realization is accurate
in a wide temperature range including the pseudocritical
temperature corresponding to the particular RF. This option (the
second) was used for the cases $\Delta=0.5$ and $1$, whereas for
$\Delta=1.5$ (and $2$~\cite{fytas06a}) both options were used,
each for $50\%$ of the simulations (for a more detailed discussion
see Ref.~\cite{fytas06a}).

\section{Results and analysis}
\label{sec:3}

\subsection{Definitions and the property of self-averaging}
\label{sec:3a}

For a disordered system we have to perform two distinct kinds of
averaging. Firstly, for each RF realization the usual thermal
average has to be carried out and secondly we have to average over
the distribution of the random parameters. In effect, this means
that we must generate and study quite large samples of RF's.
Following the methods outlined in Sect.~\ref{sec:2c} the thermal
average for the specific heat is given by the approximations
(\ref{eq:6})-(\ref{eq:8}). Using large samples of RF's we can
estimate the relevant probability distributions of the location of
the specific heat peaks. Let $C_{m}(T)$ denote the specific heat
of a particular realization $m$ in a sample of $M$ realizations of
the RF $(m=1,2,...,M)$. The pseudocritical temperature
$T_{L}^{\ast}(C_{m}(T))$ depends on the realization of the RF and
for large values of the randomness $\Delta$, is a strongly
fluctuating quantity~\cite{fytas06a}. Let us also denote the
locations of the specific heat peaks by
$(C_{m}^{\ast},T_{L,m}^{\ast})$. It seems that, in all previous
studies~\cite{hartmann01,rieger95,rieger93}, the averaging process
over an ensemble of RF's was carried out on the curve of the
averaged specific heat, without raising the question of whether
this averaged curve is the proper statistical representative of
the system. The peak of this averaged curve was then analyzed by
using finite-size scaling relations. On the other hand, the work
of Barber and Belanger~\cite{barber01}, in which the behavior was
observed for particular RF realizations, is a different route but
it would be hard to accept this as an adequate alternative. A
particular RF is not generally representative of the behavior of a
large sample of RF's.

Indeed, in previous
papers~\cite{hartmann01,rieger95,rieger93,nowak98} the following
average has been considered for the specific heat:
\begin{equation}
\label{eq:9} [C]_{av}=\frac{1}{M}\sum_{m=1}^{M}C_{m}(T)
\end{equation}
and the finite-size scaling behavior of the peak of this averaged
curve has been studied, by assuming that the maximum
$[C]^{\ast}_{av}=\max_{T}\{[C]_{av}\}$ and the corresponding
pseudocritical temperature obey the scaling laws:
\begin{equation}
\label{eq:10} [C]_{av}^{\ast}\cong p+cL^{\alpha/\nu};\;\;
T_{L}^{\ast}([C]_{av})\cong T_{c}+bL^{-1/\nu},
\end{equation}
where $\alpha$ and $\nu$ are considered to be the specific heat
and correlation length critical exponents, respectively. Note
that, these averaged curves are very sensitive to the property of
lack of self-averaging (see the discussion below) due to the fact
that the corresponding thermodynamic quantities are characterized
by broad distributions in the thermodynamic limit~\cite{fytas06a}.

It is clear that when studying random systems the only meaningful
objects for investigating the finite-size scaling behavior are the
distributions of various properties in ensembles of several
realizations of the randomness. Hence, it is important to be able
to ascertain to what extent are the results obtained from an
ensemble of random realizations representative of the general
class to which the system belongs. The answer hinges on the
important issue of self-averaging. In a self-averaging system, a
single very large system suffices to represent the ensemble;
without self-averaging, a measurement performed in a single
sample, no matter how large, does not give a meaningful result and
must be repeated on many samples. In a Monte Carlo study of a
self-averaging disordered system the number of samples needed to
obtain the average $[Q]$ (e.g., $Q$ can be the energy,
magnetization, specific heat, or susceptibility) to a given
relative accuracy decreases with increasing $L$. On the other
hand, in a non self-averaging system, the number of samples that
must be simulated rises very strongly with $L$. If a quantity is
not self-averaging, then we talk about lack of self-averaging and
as explained the process of increasing $L$ does not improve the
statistics. In other words, the sample-to-sample fluctuations
remain large. The problem of self-averaging in the RFIM has been a
matter of investigation over the last
years~\cite{fytas06a,fytas06b,parisi02}. A common measure
characterizing the self-averaging property of a system based on
the theory of finite-size scaling has been discussed by
Binder~\cite{binder88} and has been used for the study of some
random systems~\cite{wisem95,aharon96}. This measure inspects the
behavior of a normalized square width quantity, defined as:
\begin{equation}
\label{eq:11} R_{Q}=\frac{V_{Q}}{[Q]^{2}},
\end{equation}
where $V_{Q}=[Q^{2}]-[Q]^{2}$ is the sample-to-sample variance of
the average $[Q]$. Here, $Q$ is used in respect of the specific
heat $C_{m}^{\ast}$. According to the
literature~\cite{binder88,wisem95,aharon96} when the ratio $R_{Q}$
tends to a constant, the system is said to be non self-averaging
and the corresponding distribution (say $P(Q)$) does not become
sharp in the thermodynamic limit.

\begin{figure}
\resizebox{1 \columnwidth}{!}{\includegraphics{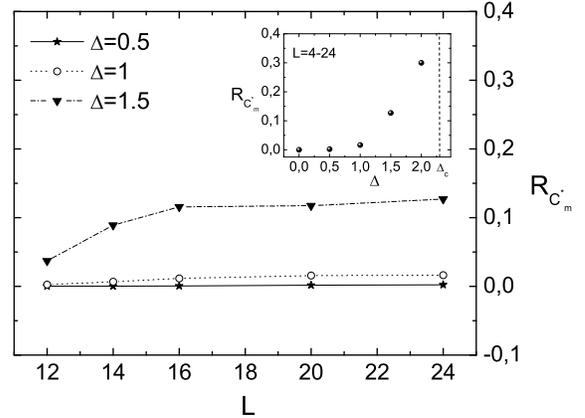}}
\caption{$L$-dependence of the ratio $R_{C_{m}^{\ast}}$ defined in
equation~(\ref{eq:11}) for $\Delta=0.5,\;1$ and $1.5$. The inset
illustrates the variation of $R_{C_{m}^{\ast}}$ as a function of
the disorder strength $\Delta$, including the value for the case
$\Delta=2$~\cite{fytas06a}.} \label{fig:1}
\end{figure}
In Ref.~\cite{fytas06a} it has been shown that the specific heat
of the bimodal RFIM for the case $\Delta=2$ is characterized by
the property of lack of self-averaging (see inset of figure 4 in
Ref.~\cite{fytas06a}). In analogy with the case $\Delta=2$ of
Ref.~\cite{fytas06a}, we construct the ratio $R_{C_{m}^{\ast}}$
for the cases $\Delta=0.5,\;1$ and $1.5$ and plot the results in
figure~\ref{fig:1}. The data presented for the cases $\Delta=0.5$
and $1$ are taken from $1000$ samples of RF's for $L\leq 10$ and
$400$ for $L=14-24$, while for the case $\Delta=1.5$ the samples
of RF's used were: $1000$ for $L\leq12$ and $300$ for $L=12-24$.

From figure~\ref{fig:1} we observe that for small values of the
randomness $\Delta$ the ratio $R_{C_{m}^{\ast}}$ has values close
to zero. In particular, for $\Delta=0.5$ the ratio
$R_{C_{m}^{\ast}}$ shows a rather faint dependence on $L$ and is
practically very close to zero. For this case, and possibly for
even weaker RFs, it appears that $R_{C_{m}^{\ast}}\rightarrow 0$
and the property of self-averaging may be well obeyed. However, we
know from our previous study for $\Delta=2$ that the above
property is strongly violated, at the same lattice sizes, and that
$R_{C_{m}^{\ast}}\rightarrow 0.3$~\cite{fytas06a}. Thus, for
strong disorder the behavior appears very different and the
development of the increasing influence of the randomness $\Delta$
on the ratio $R_{C_{m}^{\ast}}$ can be seen in figure~\ref{fig:1}
and in particular from the corresponding inset. For the lattice
sizes studied here ($L=4-24$) the ratio $R_{C_{m}^{\ast}}$ for the
cases $\Delta\geq 1$ seems to increase with the lattice size and
the estimated non-zero limiting values for $\Delta=1$ and $1.5$
are $R_{C_{m}^{\ast}}\rightarrow 0.016$ and
$R_{C_{m}^{\ast}}\rightarrow 0.12$, respectively. However, the
behavior of the specific heat is notoriously difficult even in
simple pure models~\cite{binder88}, and for the present model the
already existing conflicting situation is an additional warning
against drawing definite conclusions at these lattice sizes. It is
quite possible that we are not yet in the regime of large enough
$L$, where simple scaling laws would be expected to hold. For the
case $\Delta=2$ we have already observed~\cite{fytas06a} that the
system appears to crossover and change behavior for sizes $L>3$2
and we have suggested in that paper that the finite-size study
should be extended to at least $L=80$ in order to have a more
convincing picture. For the strengths studied here,
$\Delta=0.5,\;1$ and $1.5$, we suspect that even much larger sizes
would be needed in order to draw definite conclusions for the true
asymptotic behavior.

\begin{figure}
\resizebox{1 \columnwidth}{!}{\includegraphics{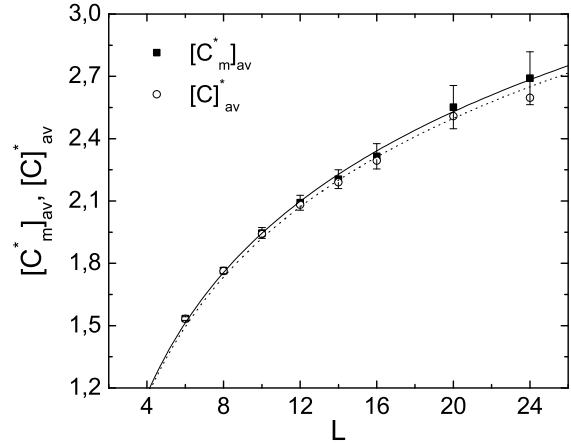}}
\caption{Finite-size behavior of $[C_{m}^{\ast}]_{av}$ and
$[C]_{av}^{\ast}$ for $\Delta=0.5$. The solid and dotted lines
correspond to logarithmic fits for $[C_{m}^{\ast}]_{av}$ and
$[C]_{av}^{\ast}$, respectively.} \label{fig:2}
\end{figure}
There are several cases in the literature where the
characterization of a phase transition demands very large linear
sizes and the picture obtained from moderate sizes is completely
misleading. A characteristic example is the 5-state 2D Potts
model, for which Landau~\cite{landa90} suggested that the expected
first-order behavior would not be clarified from finite-size data
up to sizes $L=2000$. In a different inquiry Hilfer \emph{et
al.}~\cite{hilfer03} estimated that the asymptotic behavior of the
tail regime of the universal order-parameter distribution for the
2D Ising model would require sizes of the order of $L\geq 10^{5}$.
Thus, having to deal with the controversial 3D RFIM, for which
even the existence a tricritical point at high fields is not yet
clarified~\cite{eich96}, we prefer to regard the observed in
figure~\ref{fig:1} strong violation of the self-averaging property
as a rather tentative conclusion which has to be further verified
by studying larger systems and more physical properties (such as
the magnetic susceptibility~\cite{fytas06b}).

Since all past finite temperature studies were attempted on small
and moderate sizes ($L\leq 20$), it is valuable to examine the
implications of the strong violation of the self-averaging
property at these moderate sizes. In our opinion the inconsistent
estimations in the literature have, at least partly, their origin
on such an unconventional behavior of the RFIM. In order, to
observe better these implications we proceed to study, in addition
to the above scaling laws, the sample averages of the individual
specific heat maxima and pseudocritical temperatures defined by:
\begin{eqnarray}
\label{eq:12}&[C_{m}^{\ast}]_{av}&\equiv \frac{1}{M}\sum_{m}
C_{m}^{\ast}\cong
\widetilde{p}+\widetilde{c}L^{\widetilde{\alpha}/\widetilde{\nu}};\nonumber\\
&[T_{L,m}^{\ast}]_{av}&\equiv \frac{1}{M}\sum_{m}
T_{L,m}^{\ast}\cong
\widetilde{T}_{c}+\widetilde{b}L^{-1/\widetilde{\nu}}.
\end{eqnarray}
The mean values defined above characterize the corresponding
probability distributions and consist a different kind of
representative of the samples of RF's. To quantify the
sample-to-sample variations of the specific heat peaks we use the
standard deviation of $C_{m}^{\ast}$ over a sample of
$m=1,2,...,M$ RF realizations, $V_{C_{m}^{\ast}}$. This is the
parameter of equation~(\ref{eq:11}) and figure~\ref{fig:1} and
will be also illustrated in the following figures as error bars.
However, it should not be in any case confused with the
statistical errors resulting from the thermal average
approximations of equations~(\ref{eq:6}) and (\ref{eq:7}).

\subsection{Scaling behavior of the specific heat}
\label{sec:3b}

\begin{figure}
\resizebox{1 \columnwidth}{!}{\includegraphics{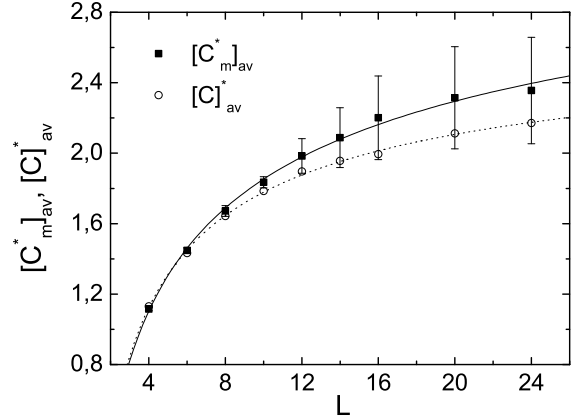}}
\caption{Finite-size behavior of $[C_{m}^{\ast}]_{av}$ and
$[C]_{av}^{\ast}$ for $\Delta=1$. The solid and dotted lines
correspond to power law fits for $[C_{m}^{\ast}]_{av}$ and
$[C]_{av}^{\ast}$, respectively. Both quantities saturate, however
with different exponents.} \label{fig:3}
\end{figure}
Let us start by presenting in figure~\ref{fig:2} the finite-size
behavior of the peaks of the sample average $[C_{m}^{\ast}]_{av}$
and that of the averaged curve $[C]_{av}^{\ast}$, for the case
$\Delta=0.5$. The number of RF realizations is $M=1000$ for $L\leq
10$ and $M=400$ for $L=12-24$. The difference between the behavior
of the peaks of the averaged curve $[C]_{av}^{\ast}$ and that of
the sample average $[C_{m}^{\ast}]_{av}$ does not emerge for small
values of the lattice size $L$. Only for $L>16$ is the
sample-to-sample fluctuation considerable and seems to
differentiate, although mildly, between the two averages.
Noteworthy that, in this case the standard deviation of the
sample-to-sample fluctuations is significantly smaller than the
average $[C_{m}^{\ast}]_{av}$:
$V_{C_{m}^{\ast}}\ll[C]_{av}^{\ast}<[C_{m}^{\ast}]_{av}$. Based on
the data $L=6-24$, no sign of saturation for both
$[C_{m}^{\ast}]_{av}$ and $[C]_{av}^{\ast}$ is observed, and one
would be tempted to predict a mildly diverging behavior. In fact
the best fits, corresponding to the smallest value of $\chi^{2}$
per degree of freedom, predict a logarithmic behavior of the form
$[C_{m}^{\ast}]_{av}=0.844(3)\cdot \ln L$ and
$[C]_{av}^{\ast}=0.834(4)\cdot \ln L$, respectively. Note that,
the fit for $[C_{m}^{\ast}]_{av}$ has a smaller value of
$\chi^{2}$ per degree of freedom than that of the fit for
$[C]_{av}^{\ast}$. Attempting a power law for
$[C_{m}^{\ast}]_{av}$ we find a diverging behavior with a much
larger value of $\chi^{2}$ per degree of freedom.

Next, we consider the intermediate case where the randomness takes
the value $\Delta=1$. Figure~\ref{fig:3} shows again the
finite-size behavior of the peaks of the sample average
$[C_{m}^{\ast}]_{av}$ and that of the averaged curve
$[C]_{av}^{\ast}$. In this case we apply power law fits of the
form: $[C]_{av}^{\ast}= p+cL^{w}$ and $[C_{m}^{\ast}]_{av}=
\widetilde{p}+\widetilde{c}L^{\widetilde{w}}$, as the one proposed
in equations~(\ref{eq:10}) and (\ref{eq:12}), using the same
number of RF realizations as in the case $\Delta=0.5$. These fits
yield saturation laws for both cases $[C_{m}^{\ast}]_{av}$ and
$[C]_{av}^{\ast}$, predicting however different saturation values
$4.15(65)$ and $2.88(13)$, respectively. Specifically, the
relevant fits of comparable $\chi^{2}$ quantity, give:
$(\widetilde{p},\widetilde{c},\widetilde{w})=(4.15(65),-4.68(44),-0.31(9))$
and $(p,c,w)=(2.88(13),-3.56(64),-0.51(56))$. From these, we could
even speculate that the saturation of both quantities takes place
with different exponents: $\widetilde{w}=-0.31$ and $w=-0.51$.
This value of $w$ corresponding to the peaks of the averaged curve
$[C]_{av}^{\ast}$ compares very well to the value $-0.5$ of
Ref.~\cite{rieger93} for the ratio $h/T=0.25$ of their study ($h$
is used for the disorder strength in Ref.~\cite{rieger93}). Using
our approximate phase diagram (see below figure~\ref{fig:6}), we
find that their case closely corresponds to the case $\Delta=1$
studied here. However, the values of the exponents estimated from
the fits can not be taken seriously, since this analysis does not
account for the systematic problem that at least part of the data
are not yet in the regime of large enough $L$, where finite-size
scaling without corrections holds. Note that, the standard
deviation of the sample-to-sample fluctuations in this case is
also smaller than $[C_{m}^{\ast}]_{av}$, that is
$V_{C_{m}^{\ast}}<[C]_{av}^{\ast}<[C_{m}^{\ast}]_{av}$, but not as
small as in the case $\Delta=0.5$.
\begin{figure}
\resizebox{1 \columnwidth}{!}{\includegraphics{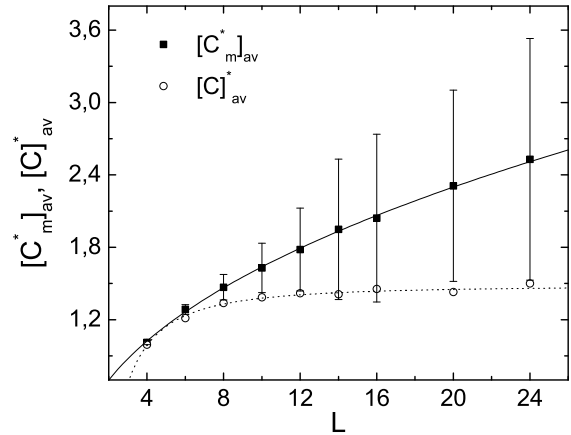}}
\caption{The same as in figure~\ref{fig:3}, but for $\Delta=1.5$.
The clear and early saturation of $[C]_{av}^{\ast}$ is similar to
that of the case $\Delta=2$~\cite{fytas06a}. The diverging power
law behavior of $[C_{m}^{\ast}]_{av}$ shown by the solid line
gives an exponent $\widetilde{w}=0.44(7)$.} \label{fig:4}
\end{figure}

Finally we treat the case $\Delta=1.5$. Figure~\ref{fig:4}
illustrates the finite-size behavior of the peaks of the sample
average $[C_{m}^{\ast}]_{av}$ and that of the averaged curve
$[C]_{av}^{\ast}$. The data presented here are taken from samples
of $M=1000$ RF's for $L\leq12$ and $M=300$ for $L=14-24$. The
saturation of the peaks for the averaged curve is quite obvious
and is attained, as in the case $\Delta=2$~\cite{fytas06a},
already in the small $L$-regime. Furthermore, the behavior of the
average $[C_{m}^{\ast}]_{av}$ looks similar with that of the case
$\Delta=2$ for $L\leq 24$, and despite the fact that there are no
signs of saturation of this quantity for these lattice sizes, the
possibility of crossing over to a final saturation for larger
lattice sizes can not be excluded. It is worth noting that, as in
the case $\Delta=2$, the standard deviation of the
sample-to-sample fluctuations seems to obey the same behavior with
that of $[C_{m}^{\ast}]_{av}$, since $V_{C_{m}^{\ast}}\sim
2([C_{m}^{\ast}]_{av}-[C]_{av}^{\ast})$ and $[C]_{av}^{\ast}\simeq
1.48$. Our power law fitting attempts predict for
$[C_{m}^{\ast}]_{av}$ a diverging behavior with
$\widetilde{p}=-0.21(3)$, $\widetilde{c}=0.7(2)$, and
$\widetilde{w}=0.44(7)$. Meanwhile, the averaged curve
$[C]_{av}^{\ast}$ strongly saturates with $p=1.48(2)$,
$c=-5.15(1.6)$ and an exponent $w=-1.69(24)$, already from the
small $L$-regime. The saturation exponent of the averaged curve
$w=\alpha/\nu=-1.69(24)$ should be compared to the value
$w=\alpha/\nu=-1.1(4)$ given in Ref.~\cite{rieger93} for
$h/T=0.5$, which now corresponds approximately to our $\Delta=1.5$
case (see figure~\ref{fig:6}).
\begin{figure}
\resizebox{1 \columnwidth}{!}{\includegraphics{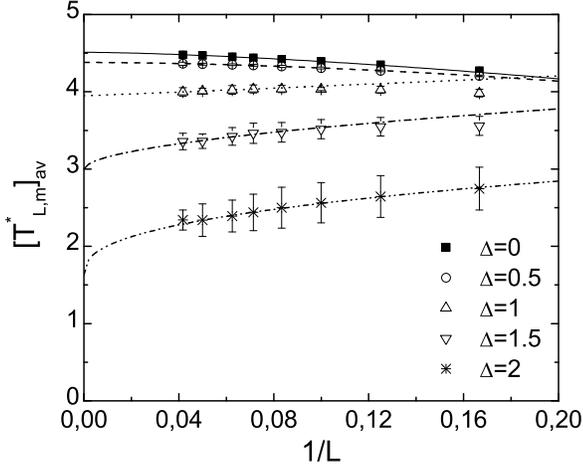}}
\caption{Size dependence of pseudocritical temperatures for
various values of $\Delta$, including the case $\Delta=0$ of the
normal cubic Ising model~\cite{malak04}, and the case
$\Delta=2$~\cite{fytas06a}.} \label{fig:5}
\end{figure}

Comparing the behavior of $[C_{m}^{\ast}]_{av}$ for the cases
$\Delta=1$ and $\Delta=1.5$, one may discern a conflicting
picture, in a sense that while $\widetilde{w}$ is negative for
$\Delta=1$ - indicating a strong saturation - the same exponent
turns out to be positive - indicating a rather strong divergence -
for $\Delta=1.5$. We believe though, that this is not a surprise.
In fact, the blowing up of the property of lack of self-averaging
in the range $\Delta=1-1.5$, as illustrated in the inset of
figure~\ref{fig:1}, may be behind this behavior. A possible
saturation in the asymptotic limit may occur in both cases but
this may happen via different complex routes because of the
unsettled and $(\Delta,L)$-sensitive self-averaging property of
the system. On the other hand, the quantity $[C]_{av}^{\ast}$ is
very weakly $L$-depended in the large $L$-regime and its early
saturation to a value (that depends on the disorder strength),
leaves no room for an accurate estimation of its behavior since
the statistical errors dominate in the large $L$-regime. This
fact, when combined with the possible crossover behavior of the
system at quite large linear sizes, larger than those
corresponding to the above discussed saturation, lead us to
suggest that scaling attempts on $[C]_{av}^{\ast}$, including
previous studies, should not be trusted.

\subsection{Phase diagram and universality aspects}
\label{sec:3c}

In figure~\ref{fig:5} we present the size dependence of the
pseudocritical temperatures for all values of randomness studied.
We have included the case of the normal cubic Ising model, for
which the numerical data of Ref.~\cite{malak04} have been used,
and the case $\Delta=2$~\cite{fytas06a}, using results up to
$L=24$, where our numerical scheme is accurate. The results of the
power law fittings applied (see equation~(\ref{eq:12})) are
presented in table~\ref{tab:1}. From table~\ref{tab:1} it is
obvious that there is a strong dependence of the shift exponent
($1/\widetilde{\nu}$) on the value of the disorder strength.
While, for relatively small values of the disorder strength
$\Delta$ the shifting of the pseudocritical temperatures follows
that of the normal Ising model, for larger values of $\Delta$ the
exponent $\widetilde{\nu}$ shows an intense variation, indicating
a possible violation of universality, in agreement with the
results of Sourlas~\cite{sourlas99}. In fact, it is known that the
only theoretical arguments supporting the existence of
universality classes in random systems are based on PRG theory and
these arguments have been intensively called into questioned for
the case of the RFIM. Equivalent studies of universality
violations have been reported also in other glassy
systems~\cite{bern96}, reenforcing the view that the concept of
universality in complex systems is not fully clarified and that
more work needs to be done towards this direction.

\begin{table}
\caption{Critical temperatures and exponents for various values of
$\Delta$, including the case $\Delta=0$ of the pure Ising
model~\cite{malak04}, and the case $\Delta=2$~\cite{fytas06a}.}
\label{tab:1}
\begin{tabular}{cccc}
\hline\hline\noalign{\smallskip}
$\Delta$&$\widetilde{T_{c}}$&$\widetilde{b}$&$\widetilde{\nu}$  \\
\noalign{\smallskip}\hline\noalign{\smallskip}
0&4.51153&-3.96(14)&0.66(6)\\
0.5&4.380(3)&-4.12(51)&0.57(24)\\
1&3.949(19)&1.40(83)&0.95(28)\\
1.5&3.001(13)&1.85(48)&1.86(34)\\
2&1.63(19)&2.28(8)&2.55(49)\\
\noalign{\smallskip}\hline\hline
\end{tabular}
\end{table}
Based on the data of table~\ref{tab:1}, we give in
figure~\ref{fig:6} an approximation of the phase diagram of the
model which is comparable with the ones given in the literature
(see i.e. Refs.~\cite{newman96,itakura01}). The dotted line shows
a power law fit of the form:
\begin{equation}
\label{eq:13} T_{c}=p+q\Delta^{r},
\end{equation}
with $p=4.5114(20)$, $q=-0.59(3)$, and $r=2.29(6)$. The value of
$p$ is very close to the value of the critical temperature of the
normal 3D Ising model ($T_{c}\approx 4.51153...$), approving to a
certain extent our fitting choice. The above power law ansantz for
the phase diagram $T_{c}=T_{c}(\Delta)$ has a clear physical
motivation, which could be compared with various functions
considered in the past~\cite{newman96,machta00}. The critical
value of the randomness is estimated to be
$\Delta_{c}=2.42\pm0.18$, close to the value $2.3\pm0.2$ of Newman
and Barkema~\cite{newman96} and the value $2.35$ of
Ogielski~\cite{ogielski86}. To get a better estimate for
$\Delta_{c}$ the system should be simulated for larger values of
$\Delta$ close enough to the critical value.

\section{Concluding remarks}
\label{sec:4}

The numerical route utilized here for the study of the RFIM
consisted of the application of the multi-range WL
algorithm~\cite{wang01} in its N-fold version~\cite{schul01},
implemented within the CrMES scheme~\cite{malak04}. We hope that
the presented combination of algorithms and techniques will be
useful in further numerical studies of this and other similarly
challenging problems, such as spin glasses.

\begin{figure}
\resizebox{1 \columnwidth}{!}{\includegraphics{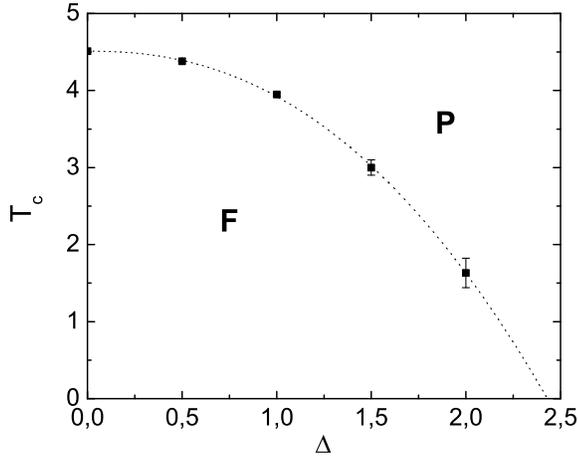}}
\caption{The phase diagram based on the data of table~\ref{tab:1}
and equation~(\ref{eq:13}). The dotted line separates the
ferromagnetic (F) from the paramagnetic (P) phase. The estimated
critical value $\Delta_{c}$ of the disorder strength, above which
no phase transition occurs, is $\Delta_{c}=2.42\pm0.12$.}
\label{fig:6}
\end{figure}
Our analysis showed that, in general, the behavior of the mean
$[C_{m}^{\ast}]_{av}$ is distinct from that of $[C]_{av}^{\ast}$
and that this is a result of the lack of self-averaging, a
property that varies strongly with the disorder strength and the
lattice sizes considered. For sufficiently small values of
randomness $\Delta$, $[C_{m}^{\ast}]_{av}$ and $[C]_{av}^{\ast}$
seem to obey a mildly diverging behavior, showing no signs of
saturation. Moving to the intermediate range $(\Delta=1)$, both
quantities saturate, but with different exponents. Yet, this area
of the phase diagram of the RFIM is not fully understood.
Theoretical studies based on the replica formalism predict the
existence of an intermediate glassy-like phase which is
characterized by a breaking of replica symmetry near the
transition temperature~\cite{belang98}. However, the
interpretation of these results is not clear, and understanding
may be simpler under the prism of a strong and complex variation
of the property of lack of self-averaging. For large values of
randomness, say $\Delta=1.5$, the peak of the averaged specific
heat curve $[C]_{av}^{\ast}$ obeys a strong saturation law which
is attained already in the small $L$-regime, in agreement with our
previous findings for an even larger value of the disorder
strength (case $\Delta=2$~\cite{fytas06a}). But as mentioned
earlier the corresponding scaling attempts would be hardly
trusted. In the same range of the disorder strength, the behavior
of the sample mean $[C_{m}^{\ast}]_{av}$ is somewhat surprising,
showing no signs of saturation, and its behavior seems to follow
the large sample-to-sample fluctuations developed by the blowing
up of the property of the lack of self-averaging. Apparently, the
blowing up of the property of lack of self-averaging in the case
$\Delta=1.5$ is responsible for this behavior, shifting a possible
saturation to larger values of $L$. Provided that our previous
analysis for the case $\Delta=2$~\cite{fytas06a} recorded a `final
and unexpected' saturating behavior of the sample average
$[C_{m}^{\ast}]_{av}$ for $L>32$, it will be interesting to
observe whether this behavior prevails for larger lattice sizes,
even in cases of small values of randomness.

Turning to the shift behavior of the pseudocritical temperatures
of the model, we found a very strong dependence of the shift
exponent on the disorder strength, reinforcing the scenario of
universality violation. The shift for small values of $\Delta$
appears to follow the direction of the pure case $(\Delta=0)$ of
shifting to $T_{c}(\Delta)$ from below and this is reflected in
the negative sign of the parameter $\widetilde{b}$. For large
values of $\Delta$, the power law exponent $\widetilde{\nu}$ shows
a very strong variation, which may be due to the existence of
additional leading and non-leading correction
terms~\cite{sourlas99}. In order to support numerically the
concept of universality for the exponent $\widetilde{\nu}$ one
should have accurate data for very large lattices, as has been
pointed out also in previous works dealing with the concept of
universality in random systems~\cite{sourlas99}. In conclusion, we
argue that the complexity of the self-averaging property for the
RFIM may be the main source behind most controversies, and we
therefore call attention to the need for studying larger systems.

\begin{acknowledgement}
This research was supported by EPEAEK/PYTHAGORAS under Grant No.
$70/3/7357$.
\end{acknowledgement}

\end{document}